\documentclass[
 reprint,
superscriptaddress,
showkeys,
 amsmath,amssymb,
 aps,
pre, 
]{revtex4-2}

\usepackage{graphicx}
\usepackage{epstopdf}
\usepackage{epsfig}
\usepackage{amssymb,amsmath,stmaryrd,tabularx}
\usepackage[utf8]{inputenc}
\usepackage[center]{subfigure}
\usepackage{scalerel,amssymb}
\usepackage{cleveref}
\usepackage[dvipsnames]{xcolor}
\usepackage[normalem]{ulem}
\usepackage{bm}
\usepackage{appendix}
\usepackage{amsmath}
\usepackage{empheq}
\usepackage{url}
\usepackage{amsmath}
\usepackage{empheq}
\usepackage{float}
\usepackage[theorems,skins]{tcolorbox}
\newtcolorbox{mymathbox}[1][]{colback=white, sharp corners, #1}
\newtcbox{\othermathbox}[1][]{nobeforeafter, math upper, tcbox raise base, enhanced, sharp corners, colback=black!10, colframe=red!30!black, drop fuzzy shadow, left=1em, top=0.5em, right=2em, bottom=0.5em}

\newcommand{\bea}{\begin{eqnarray}}  
\newcommand{\eea}{\end{eqnarray}}

\begin{document}

\preprint{APS/123-QED}
\title{Nucleation and kinematics of vortices in stirred Bose Einstein condensates}
 
\author{Jonas R\o nning and Luiza Angheluta}

\affiliation{The Njord Center, Department of Physics, University of Oslo, Blindern, 0316 Oslo, Norway\\}

\date{\today}

\begin{abstract}
We apply the Halperin-Mazenco formalism within the Gross-Pitaevskii theory to characterise the kinematics and nucleation of quantum vortices in a two-dimensional stirred Bose Einstein condensate. We introduce a smooth defect density field measuring the superfluid vorticity and is a topologically conserved quantity.  We use this defect density field and its associated current density to study the precursory pattern formations that occur inside the repulsive potential of an obstacle and determine the onset of vortex nucleation and shedding. We demonstrate that phase slips form inside hard potentials even in the absence of vortex nucleation, whereas for soft potentials they occur only above a critical stirring velocity leading to vortex nucleation. The Halperin-Mazenco formalism provides an elegant and accurate method of deriving the point vortex dynamic directly from the Gross-Pitaevskii equation. 
\end{abstract}

\pacs{Valid PACS appear here} 
\keywords{vortices, superfluid, Bose-Einstein condensat}
\maketitle

\section{Introduction}

Topological defects are the fingerprints of broken continuous symmetries, and are widely encountered in ordered systems, such as disclinations in liquid crystals~\cite{GennesP.G.de1993Tpol,poulin1997novel}, dislocations in solid crystals \cite{frank1949influence,skaugen2018dislocation,skogvollPhaseFieldCrystal2022}, orientational defects in biological active matter~\cite{thampi2014vorticity,doostmohammadi2017onset,Amiri_2022} or quantized vortices in quantum fluids~\cite{vinen1961detection,khalatnikov1965super,skaugen2016vortex}, or cosmic strings~\cite{durrer2002cosmic}. The formation and dynamics of topological defects during phase ordering kinetics through temperature quenches from the disordered phase have been well studied for decades~\cite{bray2002theory}. More recent work has been focused on developing theoretical frameworks to study non-equilibrium pattern formations and dynamical regimes of ordered systems from the collective behavior of topological defects beyond the phase-ordering kinetics.

The topological defects in an atomic Bose-Einstein condensate (BEC) are quantum vortices where the condensate is locally melted while loosing its phase coherence, and this induces persistent superfluid vortical currents outside the vortex cores. Hence, vortices are the sole carrier of circulation in the condensate~\cite{kevrekidis2007emergent}. There is active research both experimentally and theoretically on understanding and tracking the non-thermal nucleation and dynamics of quantized vortices in non-equilibrium Bose Einstein condensates. Two main frameworks are currently applied to study the creation of vortices either by rotating the condensate ~\cite{abo2001observation,fetter2001vortices,fetter2009rotating} or by coupling the condensate with a moving obstacle~\cite{raman1999evidence,onofrio2000observation,weimer2015critical,kwon2015critical,neely2010observation}. The nucleation criterion is based on the energetic argument that the superfluid flow reaches a critical velocity above which the condensate phase gradient undergoes phase slips. In rotating BEC systems, vortices are created when the system spins at a uniform frequency above a critical threshold determined by the quantized circulation of the vortex. Vortices of same circulation form at the edge of the condensate and migrate into the bulk where they eventually form vortex latices~\cite{abo2001observation,coddington2003observation,butts1999predicted,kavoulakis2000weakly,penckwitt2002nucleation}.

The vortex nucleation in condensates stirred by a moving obstacle has also been studied ~\cite{josserand1995cavitation,newell1995phase,crescimanno2000analytical,aioi2011controlled,kunimi2015metastability} and observed experimentally~\cite{neely2010observation,kwon2015critical}. Here, the nucleation criterion relies on the height $U_0$ of the repulsive potential representing the coupling of the stirring obstacle to the condensate. A hard potential corresponds to an almost impenetrable obstacle when $U_0>\mu$, where $\mu$ is the condensate chemical potential, such that the condensate density rapidly decreases and nearly vanishes inside the potential. By contrast, a soft potential corresponds to a penetrable obstacle for $U_0<\mu$ such that 
the condensate density is gently depleted inside the obstacle. The onset of vortex nucleation induced by a hard obstacle occurs when the local condensate velocity reaches the critical velocity for phonon emission, whereas for the soft obstacle this is a necessary, but not a sufficient requirement~\cite{josserand1995cavitation,newell1995phase}.
Stirring obstacles are typically modelled as Gaussian potentials with varying height and width~\cite{aioi2011controlled,kunimi2015metastability},
which approach a Dirac-delta function in the limit of a solid obstacle. 
In Ref.~\cite{kunimi2015metastability}, the vortex nucleation induced by a repulsive Gaussian potential of different strengths is studied numerically. It is found that near the critical velocity for vortex nucleation, the energy gap between the ground state and the exited state goes to zero as a power-law, while ghost vortices, i.e. phase slips, are formed inside the potential. By contrast, no such ghost vortices develop in the case of soft potentials. In addition to tuning the degree of permeability of the obstacle, different vortex shedding regimes, from vortex dipoles, pairs and clusters~\cite{sasaki2010benard,reeves2015identifying, kwon2016observation}, can be induced by varying the size of the obstacle through the width of the potential which also changes the critical stirring velocity~\cite{kwon2015critical, kwak2022minimum}. 
Once vortices are being shed into the condensate, they interact with each other forming dynamic clusters that sustain energy cascades and two-dimensional quantum turbulence~\cite{kwon2016observation,neely2013characteristics,billam2014onsager,skaugen2016vortex,Reeves_2013,Bradley_2012}. 

Even through compressibility effects, due to shock waves and phonons, are particularly important in the nucleation and the annihilation of vortex dipoles, they are typically overlooked in the quantum turbulence regime where turbulent energy spectra and clustering behavior is attributed mostly to the mutual interactions between vortices~\cite{yu2017emergent,billam2014onsager}. The point vortex modelling approach have been employed to characterised quantum turbulence from the dynamics of the point vortices~\cite{hess1967angular,middelkamp2011guiding,navarro2013dynamics,skaugen2016velocity,groszek2018motion}. In point vortex models, vortices are reduced to charged point particles with an overdamped dynamics where their velocity is determined by the mutual interaction potential or external potentials. For finite domains, boundary conditions are satisfied by adding mirror vortices to the interaction potential~\cite{NewtonPaulK2001TNP:,groszek2018motion}. 

An accurate, non-perturbative method of deriving the velocity of topological defects directly from the evolution of the order parameter of the $O(2)$ broken rotational symmetry has been developed by Halperin and Mazenko~\cite{halperin1981published,mazenko1997vortex,mazenko2001defect}. Topological defects are located as zeros in the 2D vector order parameter, where the magnitude of order vanishes to regularise the region where the phase of the order parameter becomes undefined. The defect velocity is determined by the magnitude of the defect density current at the defect position. In the frozen phase approximation, where the phase of the order parameter is stationary apart from its moving singularities, the vortex kinematics determined by the evolution of the order parameter reduces to a point vortex model~\cite{mazenko1997vortex, groszek2018motion}. Within the Gross-Pitaevskii theory, the order parameter is the condensate wavefunction and the frozen-phase approximation is the regime where the dynamics of phonon modes can be neglected. This is a versatile formalism which has been applied to various systems from  point dislocations~\cite{skaugen2018dislocation} and dislocation lines ~\cite{skogvollPhaseFieldCrystal2022} in crystals, to point orientational defects in active nematics~\cite{angheluta2021role} and active polar systems~\cite{andersen2022symmetry} and disclination lines in nematic liquid crystals~\cite{schimming2022singularity}.

In this paper, we adopt the Halperin-Mazenko formalism to gain further theoretical insights into the kinematic and nucleation of vortices in a stirred BEC that is described by the Gross Pitaevskii equation ~\cite{Bradley_2012,Reeves_2013, skaugen2016vortex}. In Section \ref{sec:MH_method}, we present the Halperin-Mazenko formalism for 2D BECs, and show that the defect density field $D$ represents a generalized, smooth vorticity, defined as the curl of the superfluid current, and its evolution determines the vortex velocity. This method circumvents the need of operating directly with the singularities in the condensate phase, which are harder to manipulate both theoretically and numerically. In Section \ref{sec:nucleation}, we study how the defect field $D$ and its corresponding current density $\mathbf J^D$ are behaving during a nucleation event for two representative stirring potentials.  In the case of a soft potential, the $D$-field shows the gradual buildup of generalized vorticity around the the edge of the potential and before any phase-slips occur. This provides us with an accurate measure of the precursory pattern leading to a nucleation event. By contrast, the pattern of the $D$-field for a hard potential is given by a uniform halo formed around the edges of the potential coexisting with phase slips inside the potential, also referred to as ghost vortices. At the onset of nucleation, the superfluid vorticity halo localises into two regions while a dipole of phase slips migrates and occupies them to form a vortex dipole which sheds from the edge of the potential. In Section~\ref{sec:kinematics}, we derive the point vortex dynamics and discuss the effect of non-uniform condensate density on the vortex velocity. Concluding remarks and a summary are presented in Section \ref{sec:discussion}.  

\begin{figure}[ht]
    \centering
    \includegraphics[width=0.45\textwidth]{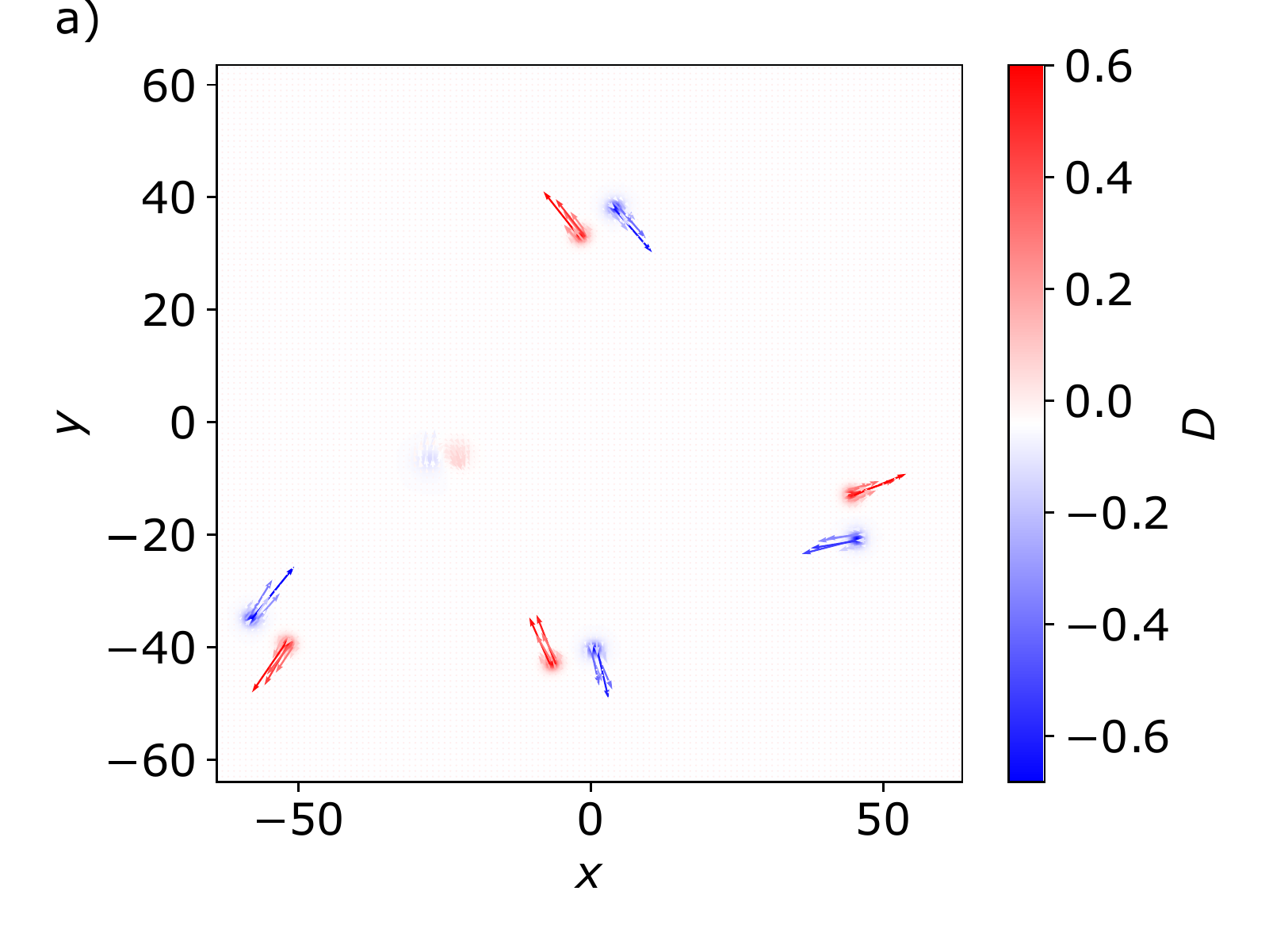}
    \includegraphics[width=0.45\textwidth]{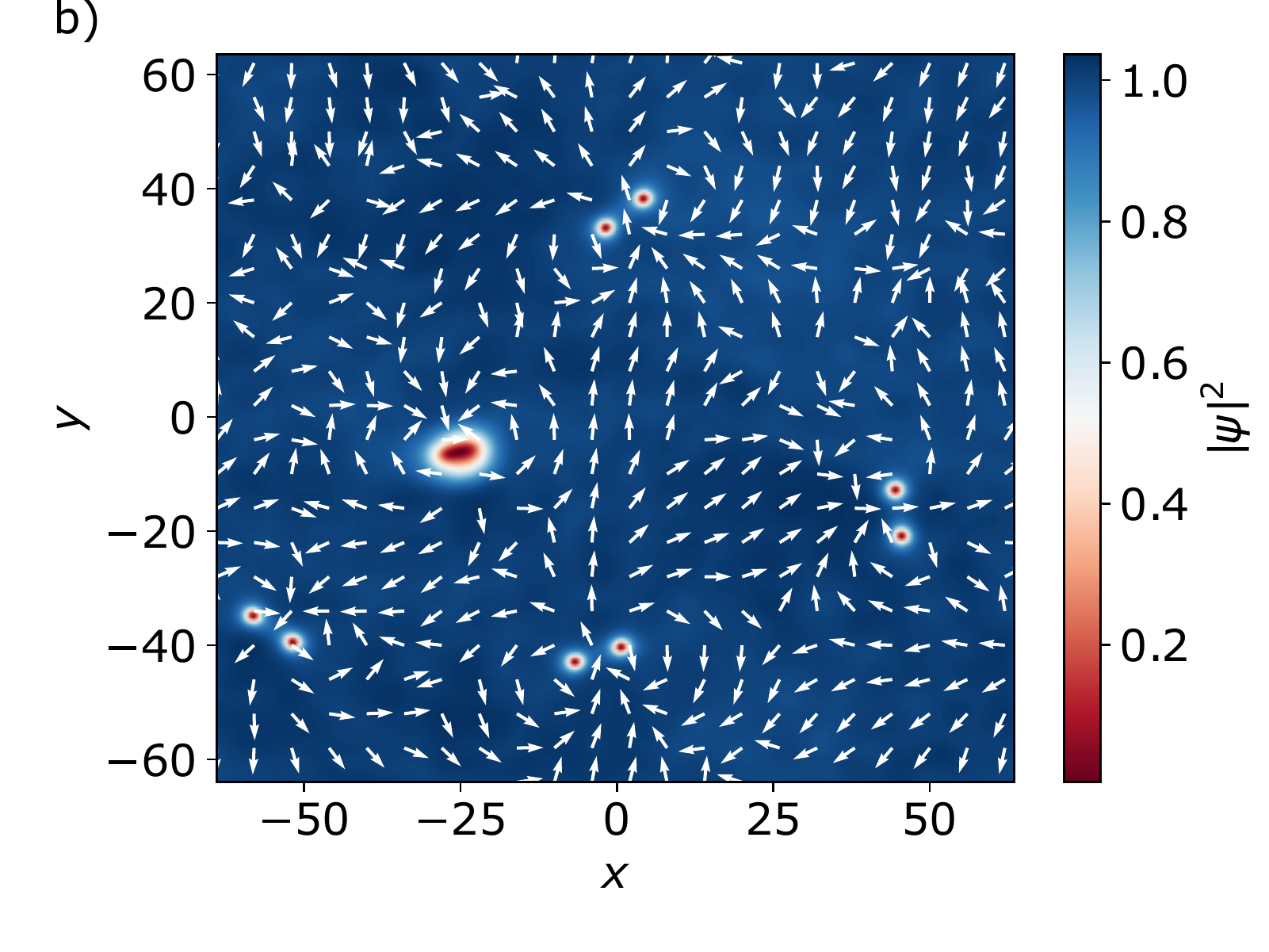}
    \caption{Snapshot of the $D$-field (a) and condensate density (b) in the presence of a stirring Gaussian potential and shed vortex dipoles. The vector-fields are the defect current $\mathbf J^D$ (a) and the normalized superfluid current $\mathcal J$ (b), respectively. Negative vortices move in the opposite direction of the defect current $\mathbf J^D$. }
    \label{fig:D-field}
\end{figure}

\section{Vortices as moving zeros}\label{sec:MH_method}
The superfluid flow and the topological structure of a weakly-interacting BEC are described by the evolution of its macroscopic wavefunction $\psi=|\psi|e^{i\theta}$, where $|\psi|$ is the condensate density and $\theta$ is the condensate phase which is coherent, i.e. constant at equilibrium. Disturbances in the condensate phase generate a superfluid flow with a current (momentum) density 
\begin{equation}\label{eq:superfluid_current}
    \mathcal J = |\psi|^2 \nabla \theta = \textrm{Im}(\psi^*\nabla\psi),
\end{equation}
such that gradients in the condensate phase define the superfluid flow velocity, which is irrotational everywhere except at the points $\mathbf r_\alpha$ where the condensate phase is undetermined (singular), namely
\begin{equation}\label{eq:curl_nabla_theta}
\nabla\times\nabla \theta = 2\pi q_\alpha \delta^2(\mathbf r-\mathbf r_\alpha). 
\end{equation}
This phase singularity has a topological nature determined by a $\pm 2\pi$ phase jump upon going counterclockwise around a loop $C_\alpha$ enclosing it. This corresponds to a quantum vortex with a quantized circulation given by the topological charge $q_\alpha = \pm 1$ determined by the  contour integral surrounding the defect 
\begin{equation}\label{eq:q_alpha_integral}
    2\pi q_\alpha = \oint\limits_{C_\alpha} d\theta = \oint\limits_{C_\alpha} d\mathbf l \cdot \nabla \theta.
\end{equation}
which is the integral form of the differential form in Eq.~(\ref{eq:curl_nabla_theta}). A singular vortex charge density field can be constructed by a superposition of the delta functions localizing vortices as phase singularities in space  
\begin{equation}
    \rho_v(\mathbf r,t) = \sum_\alpha q_\alpha \delta^2(\mathbf r -\mathbf r^{(\alpha)}(t)),
    \label{eq:vortex_density}
\end{equation}
and which represents the singular vorticity field as the curl of the superfluid flow velocity. Since, the wavefunction is well-defined everywhere including at the vortex position, it means that the condensate density vanishes where the condensate phase is undetermined. This implies that both the real and the imaginary parts must vanish. By representing the complex wavefunction into a $O(2)$-symmetric real vector field $\Vec \Psi = [\Psi_1 ; \Psi_2 ]$, 
where $\Psi_1 =\textrm{Re}(\psi)$ and $\Psi_2 = \textrm{Im}(\psi)$, we notice that $\vec \Psi(\mathbf r)$ maps a point $\mathbf r$ to a point in the $(\Psi_1,\Psi_2)$-disk centered at origin and of unit radius (i.e. $|\psi_0|^2=1$ in rescaled units). Regions of uniform condensate density map to the unit circle, whereas vortices located at various positions $\mathbf r_\alpha$ in the real space reside at the origin of the   $(\Psi_1,\Psi_2)$-disk. The coordinate transformation between the real $(x,y)$-space to the $(\Psi_1,\Psi_2)$-disk is determined by the Jacobi determinant 
\begin{equation}\label{eq:D_field}
    D = \begin{vmatrix}
     \partial_x \Psi_1 & \partial_x \Psi_2 \\
     \partial_y \Psi_1 & \partial_y \Psi_2
    \end{vmatrix}
    = \epsilon_{ij}\partial_i\Psi_1 \partial_j\Psi_2 = \frac{\epsilon_{ij}}{2i}\partial_i\psi^*\partial_j\psi,
\end{equation}
and is a scalar field that depends on the spatial position $\mathbf r$. Namely, $D(\mathbf r)$ vanishes in regions of uniform condensate and is non-zero otherwise, as it is the case around vortices. By a coordinate transformation of the Dirac delta function in Eq.~(\ref{eq:vortex_density}), we can rewrite the singular vortex density in terms of the zeros of the $\vec \Psi$ as
\begin{equation}\label{eq:q_alpha_integral2}
 \rho_v(\mathbf r,t) = D(\mathbf r,t) \delta^2(\vec \Psi).
\end{equation}

Figure~\ref{fig:D-field} (a) shows the colormap of the $D$-field identifying the vortex positions and their charges. The vector field corresponds to the current of the $D$-field as a conserved quantity, as discussed later. It turns out that the $D$-field is a measure of the non-singular superfluid vorticity given by the curl of the superfluid current from Eq.~(\ref{eq:superfluid_current}), namely
\begin{equation}\label{eq:D_J}
    \epsilon_{ij}\partial_j \mathcal J_i =\epsilon_{ij}\textrm{Im}(\partial_i\psi^*\partial_j\psi)= 2D.
\end{equation}
Whilst the topological structure in the condensate phase has a singular nature, vortices in the condensate have a non-zero core and are described by a smooth generalized vorticity field identified with the $D$-field. To show that the $D$-field indeed captures all topological content of the condensate phase, we integrate Eq.~(\ref{eq:D_J}) over an area $S$ containing vortices
\begin{equation}\label{eq:integral_relation}
    \int\limits_S d^2\mathbf r D  = \frac{1}{2} \int\limits_{\partial S} d\mathbf l \cdot \mathcal J = \frac{1}{2} \int\limits_{\partial S} |\psi|^2 \nabla \theta \cdot d\mathbf l. 
\end{equation}
In the assumption that the contour $\partial S$  enclosing the area $S$ is sufficiently far away from well-separated vortices, i.e. the superfluid density equals its uniform bulk value $|\psi_0|^2 = 1$ along the integration contour, the above integral reduces to 
\begin{equation}
     \int\limits_S d^2\mathbf r D  = \pi \sum_{\alpha \in S} q_\alpha,
     \label{eq:integral_D_const_rho}
\end{equation}
where the sum is over all vortices inside the contour. Equivalently, integrating the absolute value of $D$, we obtain instead the total number $N$ of vortices enclosed by the contour, 
\begin{equation}\label{eq:abs_D_int}
    \int\limits_{ S} d^2\mathbf r |D| = \sum_\alpha q_\alpha\int\limits_{S_\alpha} d\mathbf r  D = N \pi. 
\end{equation}
This is a limit case of a uniform condensate punctuated by well-separated vortices. However, in this formalism, the generalized vorticity $D$ field picks up not only topological defects, but any phase gradient (flow) disturbances modulated by the superfluid density, which may be induced by compressible modes, trapping or stirring potentials. These non-singular contributions become particularly important for the nucleation of vortices as discussed in Section~\ref{sec:nucleation}.

Flow disturbances with a topological origin can be dissociated from the rest by the value of the generalized vorticity determined by the condensate density profile  near a vortex. Namely, if we consider Eq.~(\ref{eq:integral_relation}) for a disc $S$ of radius much smaller that the coherence length and centered at an isolated vortex in an otherwise homogeneous condensate, we find that the value of the $D$-field reaches in magnitude a value given by
\begin{equation}
    |D_0| = \Lambda^2 \approx 0.7,
\end{equation}
using the near vortex profile of the condensate $|\psi|(r) = \Lambda r$~\cite{Pismen}, and the numerical value for the steepness $\Lambda$ of the density gradients  taken from Ref.~\cite{bradley2012energy}.

The generalized vorticity is a topologically conserved quantity as follows directly by time differentiation of Eq.~(\ref{eq:D_field}) which leads to the conservation law~\cite{mazenko1997vortex}
\begin{eqnarray}
    \partial_t D =  -\partial_i J_i^{D},
     \label{eq:timedep_D}
\end{eqnarray}
with its corresponding generalised vorticity current 
\begin{equation}
J^{(D)}_i = \epsilon_{ij}\textrm{Im}(\partial_t\psi\partial_j\psi^*),
\end{equation}
determined uniquely by the evolution of the condensate wavefunction. This vector field is also plotted in Fig.~\ref{fig:D-field} (a) where we notice that the generalized vorticity current is non-zero in regions of non-uniform superfluid flow Fig.~\ref{fig:D-field} (b), and particularly through the vortex cores where there are phase slips. The reason that the $D$-field is topologically conserved is because it determines the conservation of the singular (topological) change density $\rho_v$ which follows from the time differentiation of  Eq.~(\ref{eq:vortex_density}) and reads as 
\begin{equation}
    \partial_t\rho_v = -\partial_i J_i^{(\rho_v)},
    \label{eq:vortex_conservation}
\end{equation}
with the corresponding singular vortex current density being
\begin{eqnarray}
    \mathbf J^{(\rho_v)}(\mathbf r,t)  &=& \mathbf J^{(D)}(\mathbf r,t)\delta^2(\vec\Psi) \nonumber\\
    &=& \sum_\alpha q_\alpha \frac{\mathbf J^{(D)}(\mathbf r_\alpha)}{D(\mathbf r_\alpha)} \delta^2(\mathbf r -\mathbf r_\alpha).
\end{eqnarray}

\begin{figure}[ht]
    \centering
    \includegraphics[width=0.45\textwidth]{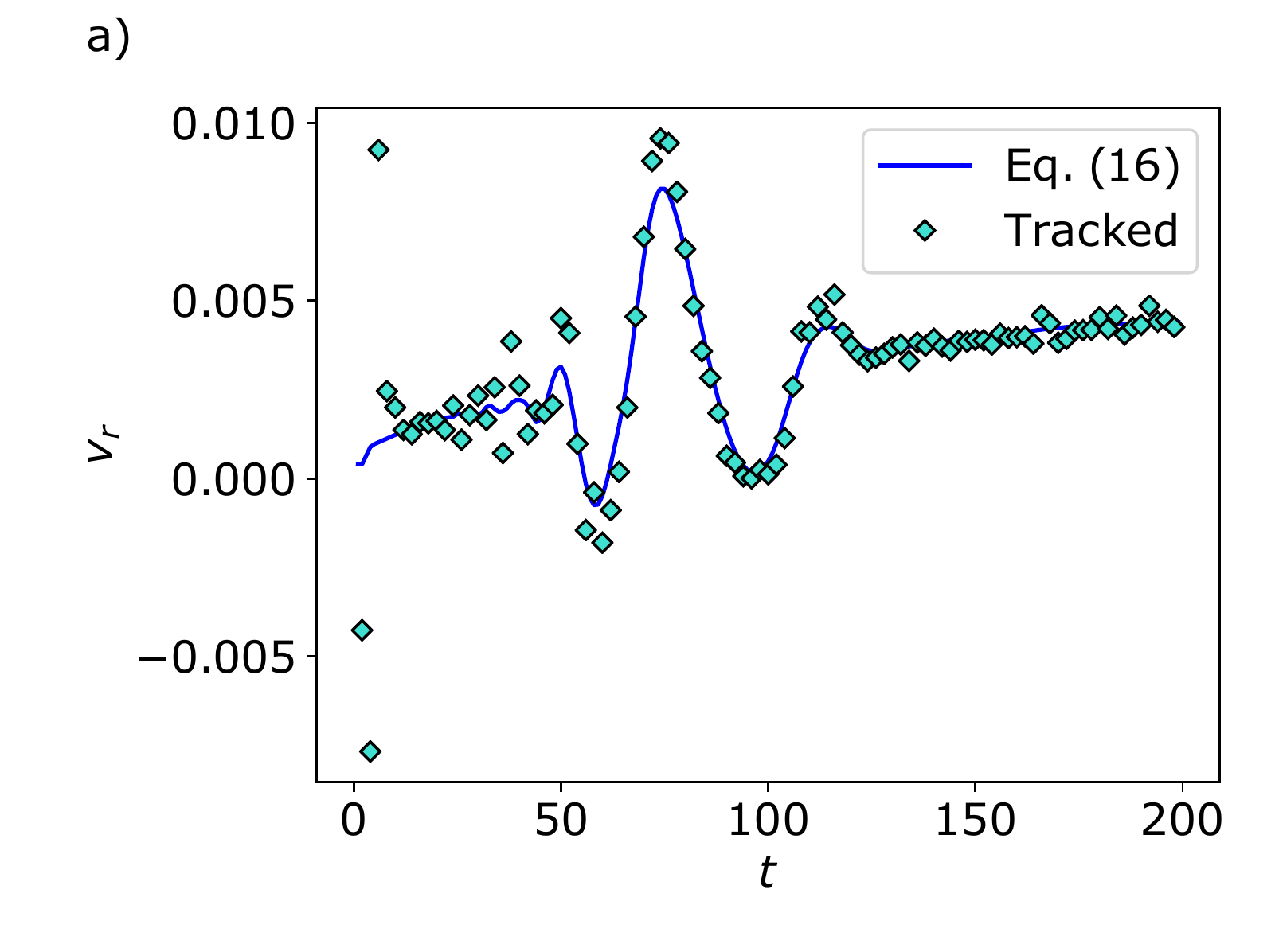}
    \includegraphics[width=0.45\textwidth]{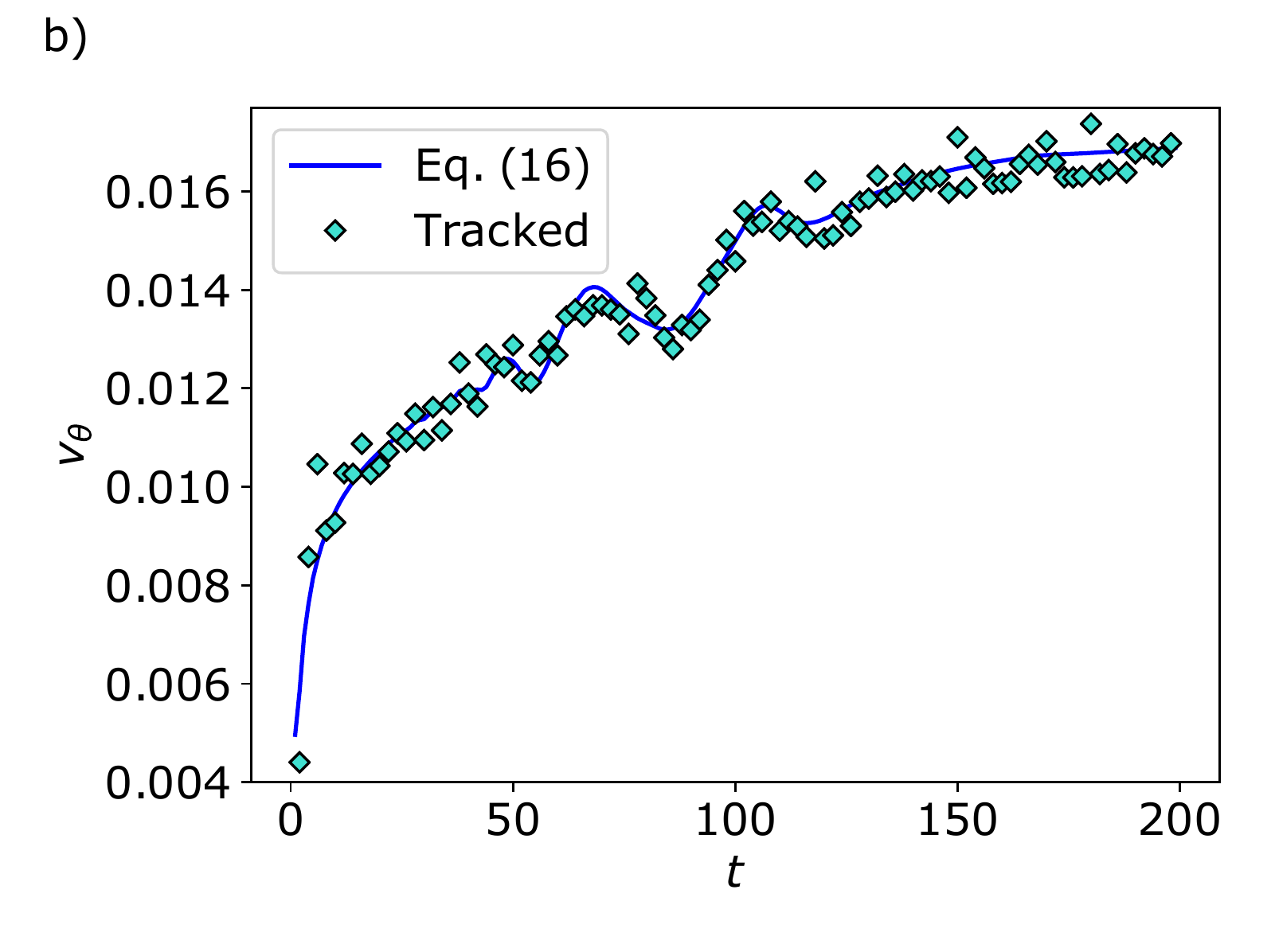}
    \caption{The radial $v_r$ (a) and angular $v_\theta$ (b) velocity component of a single vortex in a harmonic potential with $R_{tf} = 60$ obtained from Eq.~(\ref{eq:vortex_velocity}) (blue line) and from tracking the position of the defect (turquoise diamonds). The thermal dissipation is $\gamma = 0.05$. The time derivatives are computed with a timestep of $dt = 1$. }
    \label{fig:mesh_vs_mazenco}
\end{figure}

In the frozen-phase approximation, the vortex core is rigid and the equilibrium vortex wavefunction profile remains stationary in the vortex co-moving frame. In this case, the vortex current is identical to the advective current $\sum_\alpha q_\alpha \mathbf v_\alpha \delta(\mathbf r -\mathbf r_\alpha)$~\cite{Skogvoll2022Topological}. Within this approximation, the velocity is uniform or slowly-varying through the vortex core and given as
\begin{equation}
    v_\alpha=\frac{J^{D}(\mathbf r_\alpha)}{D(\mathbf r_\alpha)}.
    \label{eq:vortex_velocity}
\end{equation}
This relation provides a smooth and accurate measurement of the vortex velocity. To illustrate this, we track the trajectory of a single vortex imprinted in a BEC with a harmonic potential. At zero temperature, the vortex moves in a orbit of constant radius around the center of the harmonic trap. However, at non-zero temperature, there is a non-zero radial velocity such that the vortex spirals out towards the edge of the trap. The angular $v_\theta$ and radial $v_r$ velocity components as functions of time are shown in Figure~\ref{fig:mesh_vs_mazenco}. We notice that velocity obtained by the slope of the vortex trajectory is a noisy signal compared to the velocity from Eq.~(\ref{eq:vortex_velocity}).

Within the Gross-Pitaevskii theory, we can also find a simplified expression for the vortex velocity as discussed next. We consider that a weakly-interacting BEC at nonzero temperature where the evolution of $\psi$ in the presence of a potential field $U(\mathbf r,t)$ containing both a static trapping potential and a time-dependent stirring potential, can be described by a damped Gross-Pitaevskii equation (dGPE) which in dimensionless units reads as~\cite{Bradley_2012,Reeves_2013, skaugen2016vortex} 
\begin{equation}\label{eq:GPE}
    \partial_t \psi = (i+\gamma) \left[\frac{1}{2}\nabla^2 \psi +(1 -U-|\psi|^2)\psi \right],
\end{equation}
where the damping coefficient $\gamma$ represents the coupling of the condensate with the thermal bath.
The equation is dimensionless by rescaling the variables using the dimensional units given by the chemical potential $\mu$, the coherence length $\xi = \hbar/\sqrt{m\mu}$ and the sound velocity $c= \mu/m$. The wavefunction is rescaled in units of $\sqrt{\mu/g}$, where $m$ is the mass of the bosons and $g$ is an effective scattering parameter for the interactions between bosons. 

By inserting Eq.~(\ref{eq:GPE}) into the conservation law of the $D$-field in  Eq.~(\ref{eq:timedep_D}), we find, after some algebraic manipulations, that the evolution of the generalized superfluid vorticity is 
\begin{multline}
    \partial_t D = - \frac{1}{2}\epsilon_{ij}\partial_i\partial_k\textrm{Re}(\partial_k\psi^*\partial_j\psi) + \frac{\epsilon_{ij}}{2} \partial_i V \partial_j |\psi|^2
    \\
    +\frac{\gamma}{2} \nabla^2 D+2\gamma D [ 1 -  V -2|\psi|^2] \\
    + \gamma  \mathcal J \cdot \nabla^\perp  V    +\frac{\gamma}{2}  \epsilon_{ij} \textrm{Im}[\partial_i\partial_k \psi \partial_j\partial_k \psi^*].
\end{multline}
where the first term on rhs comes from the kinetic energy, while the second term corresponds to the coupling with an external potential and, this term is nonzero when the gradient in the condensate density is normal the gradient force. The remaining terms relate the effect of thermal damping to different sources of dissipation of superfluid vorticity, such as diffusion,  sink/sources of vorticity from the coupling with a potential $V$ and a thermal drag induced by superfluid flow. 

Since the condensate density vanishes at the vortex position, the only non-zero contribution to the generalized vorticity current density at the vortex position comes from the kinetic energy, thus the vortex velocity reduces to 
\begin{equation}\label{eq:v_alpha_general}
    \mathbf{v}_\alpha = i \frac{\textrm{Re}(\nabla^2\psi^* \nabla^\perp \psi) +\gamma \textrm{Im}(\nabla^2\psi \nabla^\perp \psi^* ) }{\nabla\psi^*\cdot\nabla^\perp\psi} \Bigg|_{\mathbf r=\mathbf r_\alpha},
\end{equation}
The contribution from the external potential $V$ plays an important role in how vortices are being nucleated and shed away from a stirring potential as discussed next. 

\section{Vortex nucleation}\label{sec:nucleation}
\begin{figure}[ht]
    \centering
    \includegraphics[width = 0.43\textwidth]{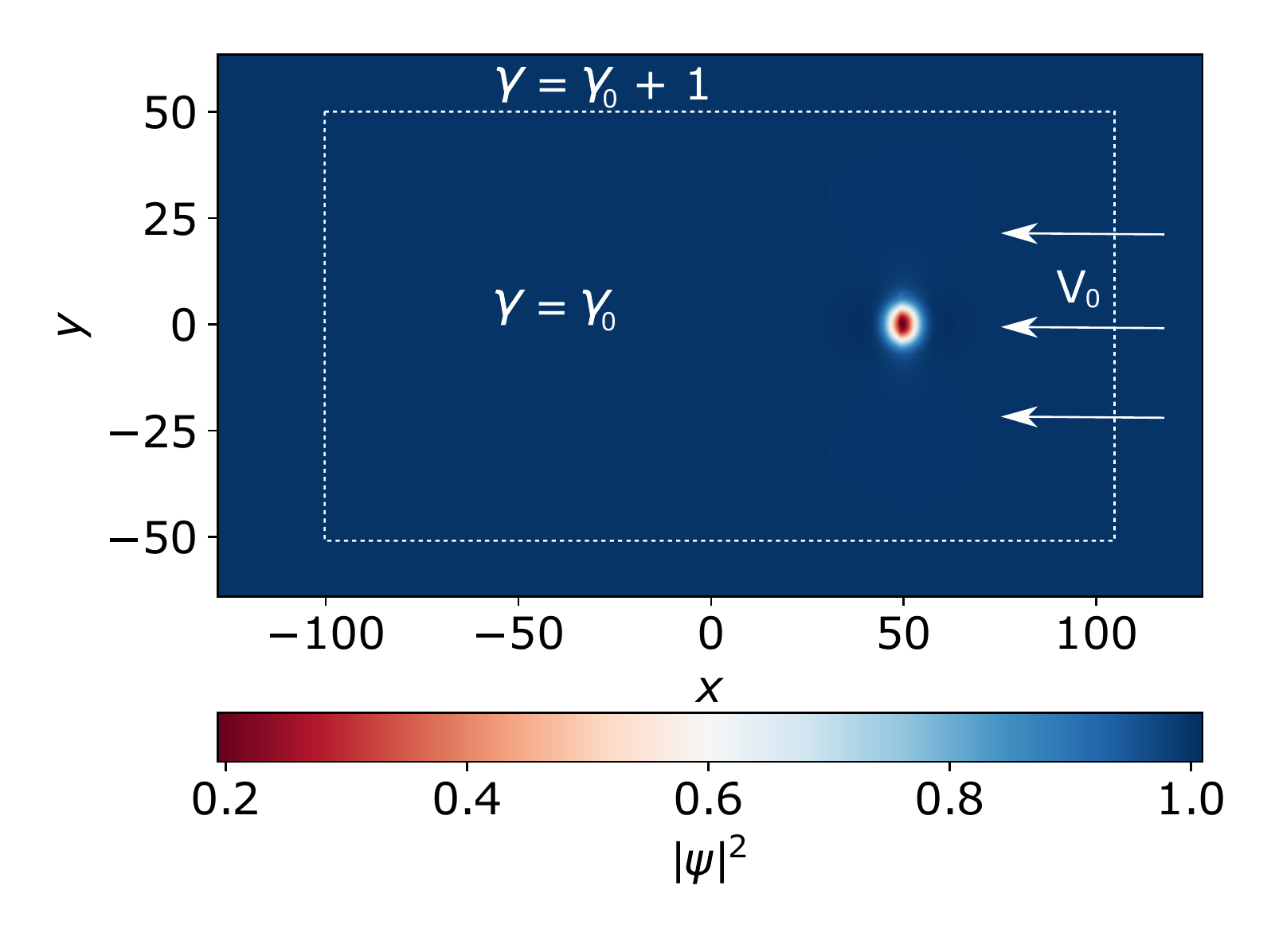}
    \caption{Numerical setup with the condensate density as a colormap. The dotted lines show the interfaces with the thermal buffer where dissipation is high. The arrows show the direction of the flow in the comoving frame. }
    \label{fig:Setuo}
\end{figure}

To study the onset of vortex nucleation, we consider a uniform Bose-Einstein condensate at zero temperature that is coupled with a Gaussian potential moving at a constant speed $V_0$ along the $x$-axis. In the co-moving frame, this is equivalent to having a static potential in a uniform superfluid flow described by   
\bea\label{eq:GPEcomoving}
    \partial_t \psi +V_0\partial_x \psi &=& (i+\gamma) \left[\frac{1}{2}\nabla^2 +(1-|\psi|^2) \right]\psi\nonumber\\
    &-& (i+\gamma)U_0 e^{-\frac{(\mathbf r -\mathbf r_0)^2}{d^2}}\psi,
\eea
where $d$ is the width of the potential and $U_0$ is the coupling strength. In numerical simulations, we consider a thermal buffer on the edge of the period domain where the damping coefficient is non-zero to avoid re-circulation of the shed vortex dipoles and to damp wave interferences. This setup is illustrated in Figure~\ref{fig:Setuo}. A similar computational trick was used in previous studies of the vortex shedding 
~\cite{reeves2015identifying} and the formation of phonon wake ~\cite{ronning2020classical}. The width of the stirring potential is set to $d =4$, and we vary its speed $V_0$ and its height $U_0$. We use a rectangular domain $[-128,128]\times[-64,64]$ (corresponding to a $512\times 256$ rectangular grid) and a fixed timestep $dt=0.01$. The potential is centered in the middle of the domain at $(x_0,y_0) = (50,0)$. For the dissipative buffer, we set the thermal drag to $\gamma(\mathbf r) = \max[\gamma_x(x),\gamma_y(y)]$, with 
\[
\gamma_x(x) =\frac{1}{2}(2 +\tanh[(x-w_x)/\Delta] -\tanh[(x+w_x)/\Delta]) +\gamma_0,
\]
and similarly for $\gamma_y(y)$. Here $\gamma_0$ is the bulk thermal drag (as presented in Figure~\ref{fig:Setuo}), $x= \pm w_x$ and $y=\pm w_y$ sets the edge of the buffer, while $\Delta$ sets the interface's width. The buffer parameters are set to $\gamma_0 =0$ (zero temperature), $w_x =100$, $w_y =50$ and $\Delta = 7$. We start by relaxing the initial Thomas-Fermi ground-state in imaginary time to find the stationary state at $V_0 =0$, and then evolve the condensate wavefunction according to Eq.~(\ref{eq:GPEcomoving}) for a given $V_0$. However, since the initial state is not the ground-state of this equation, there will be an initial disturbance forming around the potential. When the critical velocity for vortex nucleation is approached from bellow, this may trigger the nucleation of a single dipole, with no additional recurrent shedding.

\begin{figure*}[ht]
    \centering
    \includegraphics[width=0.43\textwidth]{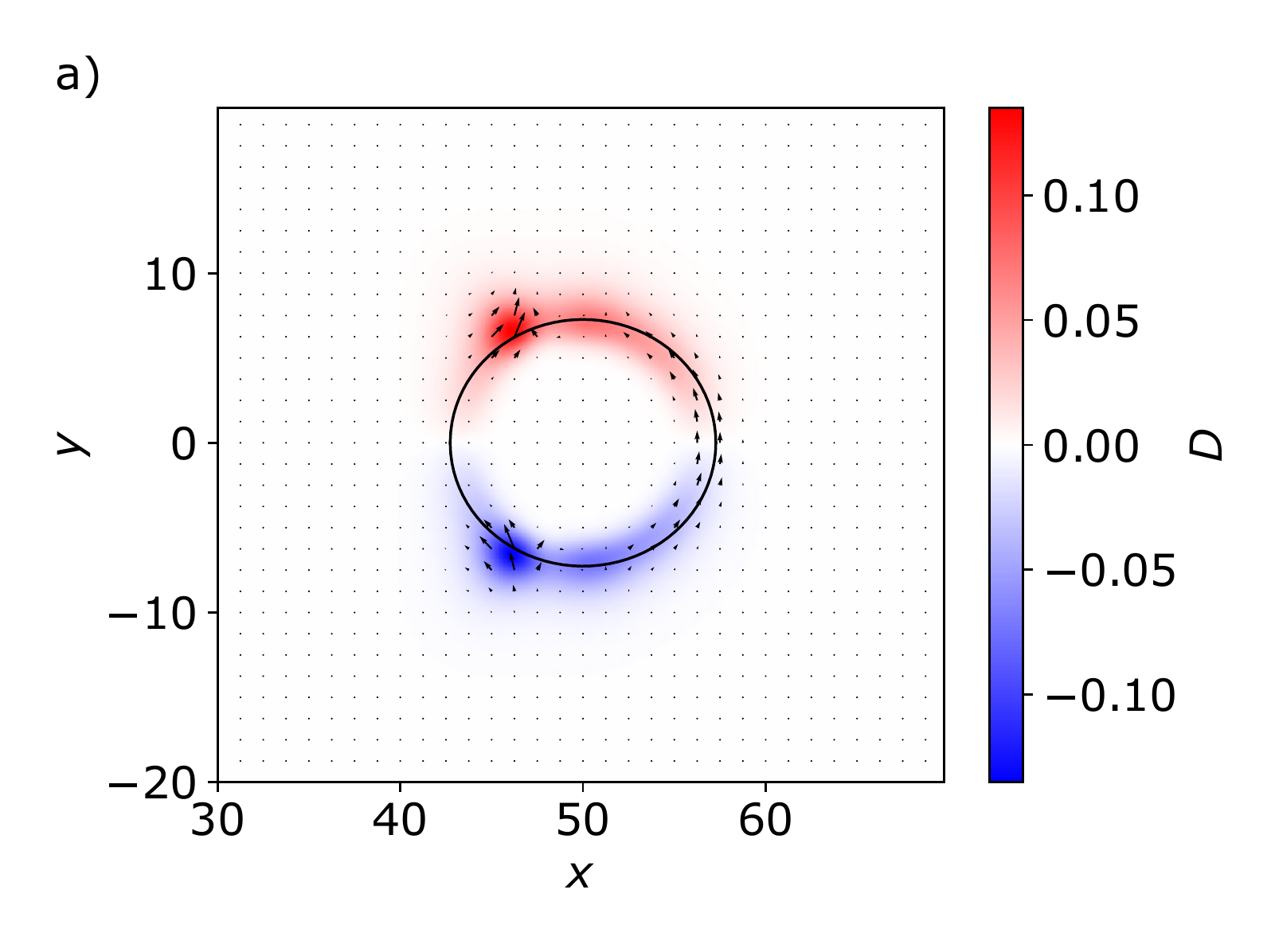}
     \includegraphics[width=0.43\textwidth]{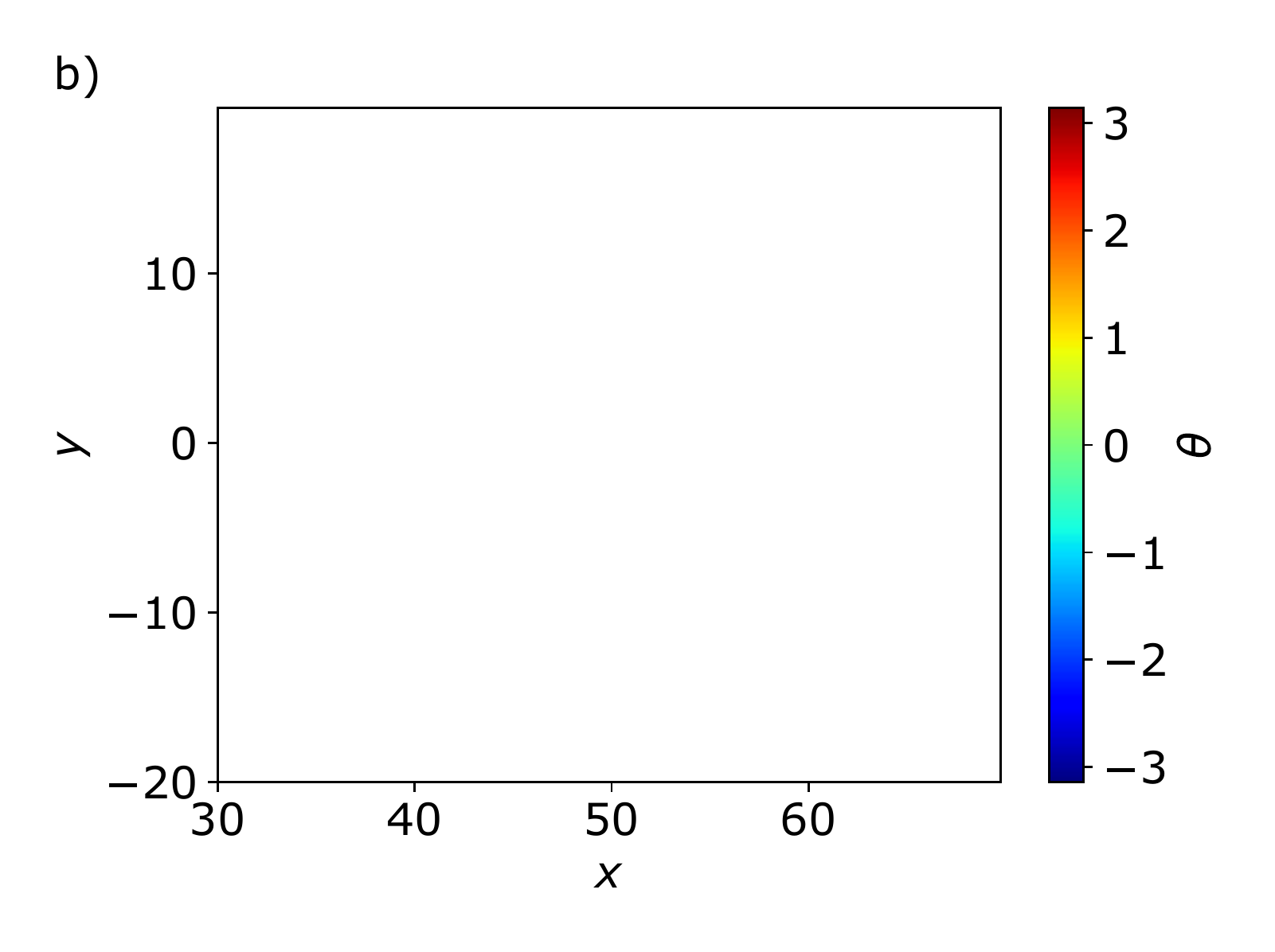}
    \includegraphics[width=0.43\textwidth]{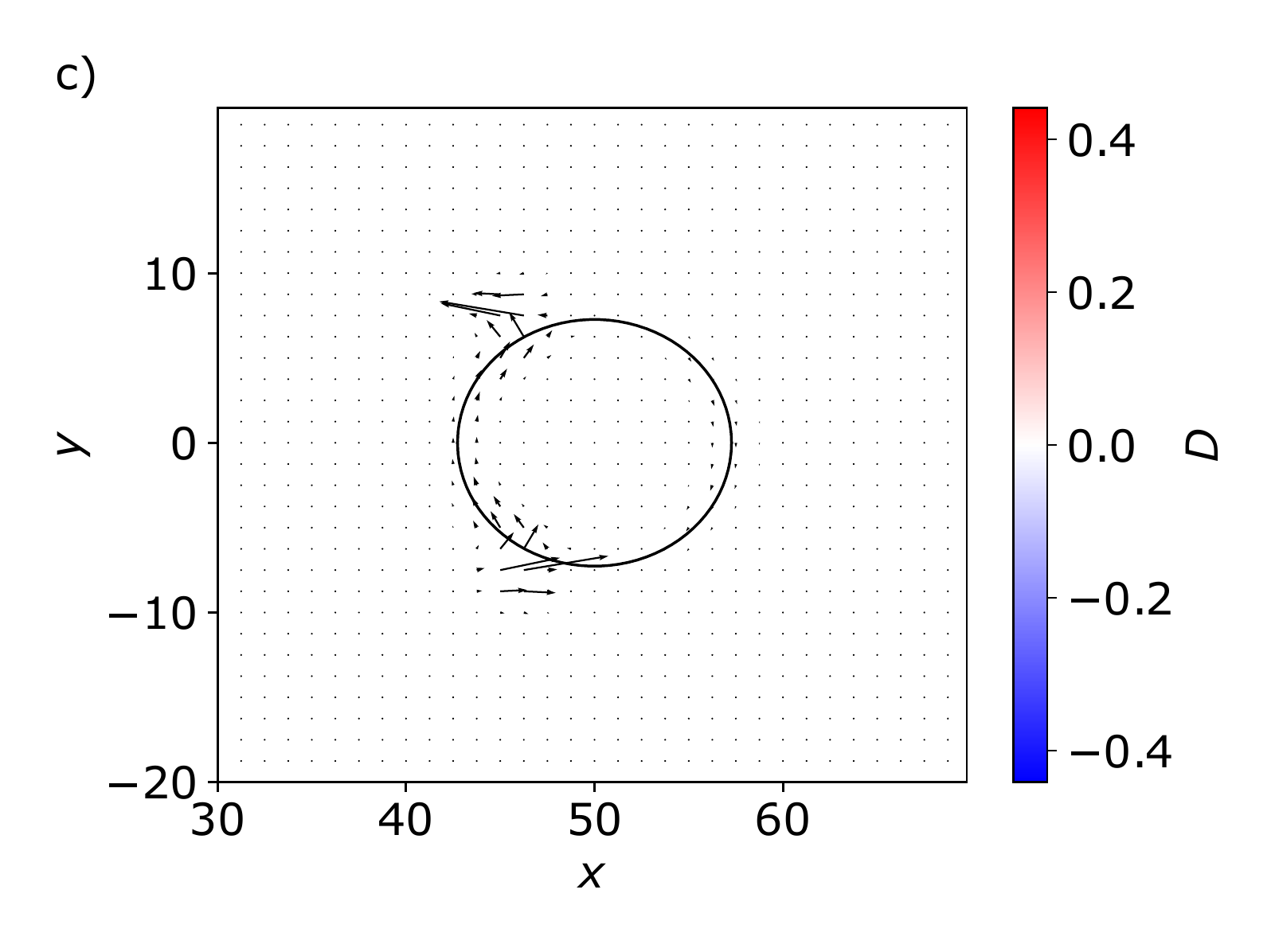}
     \includegraphics[width=0.43\textwidth]{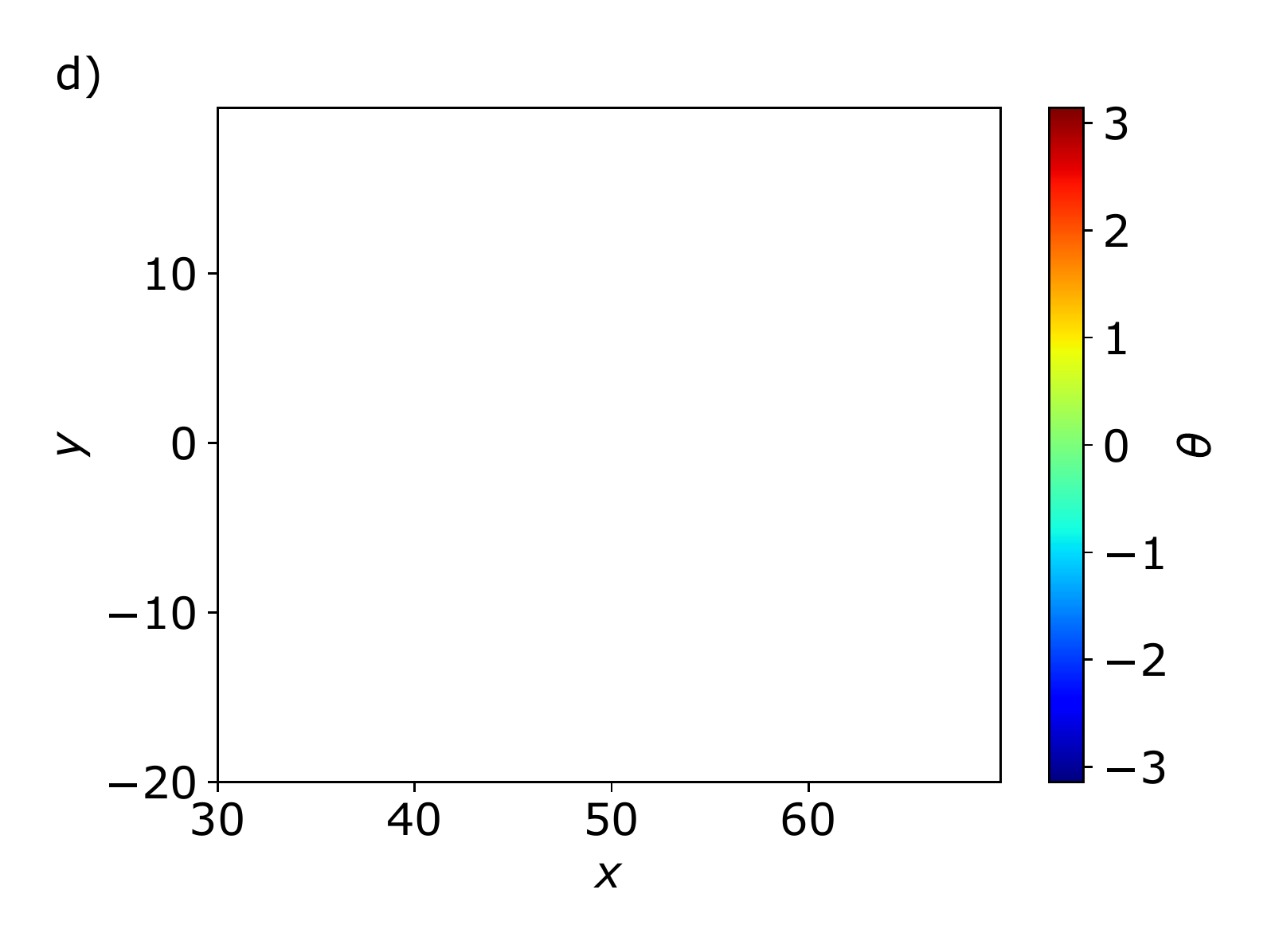}
    \caption{(a) $D$-field and (b) condensate phase $\theta$ in the vicinity of a hard potential just before the vortex dipole nucleation at $t =25$. The phase-slips reside inside the potential where the condensate density is close to zero, and the $D$ field develops a rim of intense vorticity on the edge of the potential. (c) $D$-field and (d) $\theta$ when the phase slip has moved toward the edge of the potential to localise the $D$-field into a dipole of ghost vortices at $t =50$. White/black circle is the radius where the potential becomes $e^{-1}$. The superimposed vector field in a) and c) represents the generalized vorticity current. 
    }
    \label{fig:hard_pot}
\end{figure*}

\begin{figure*}[ht]
    \centering
    \includegraphics[width=0.43\textwidth]{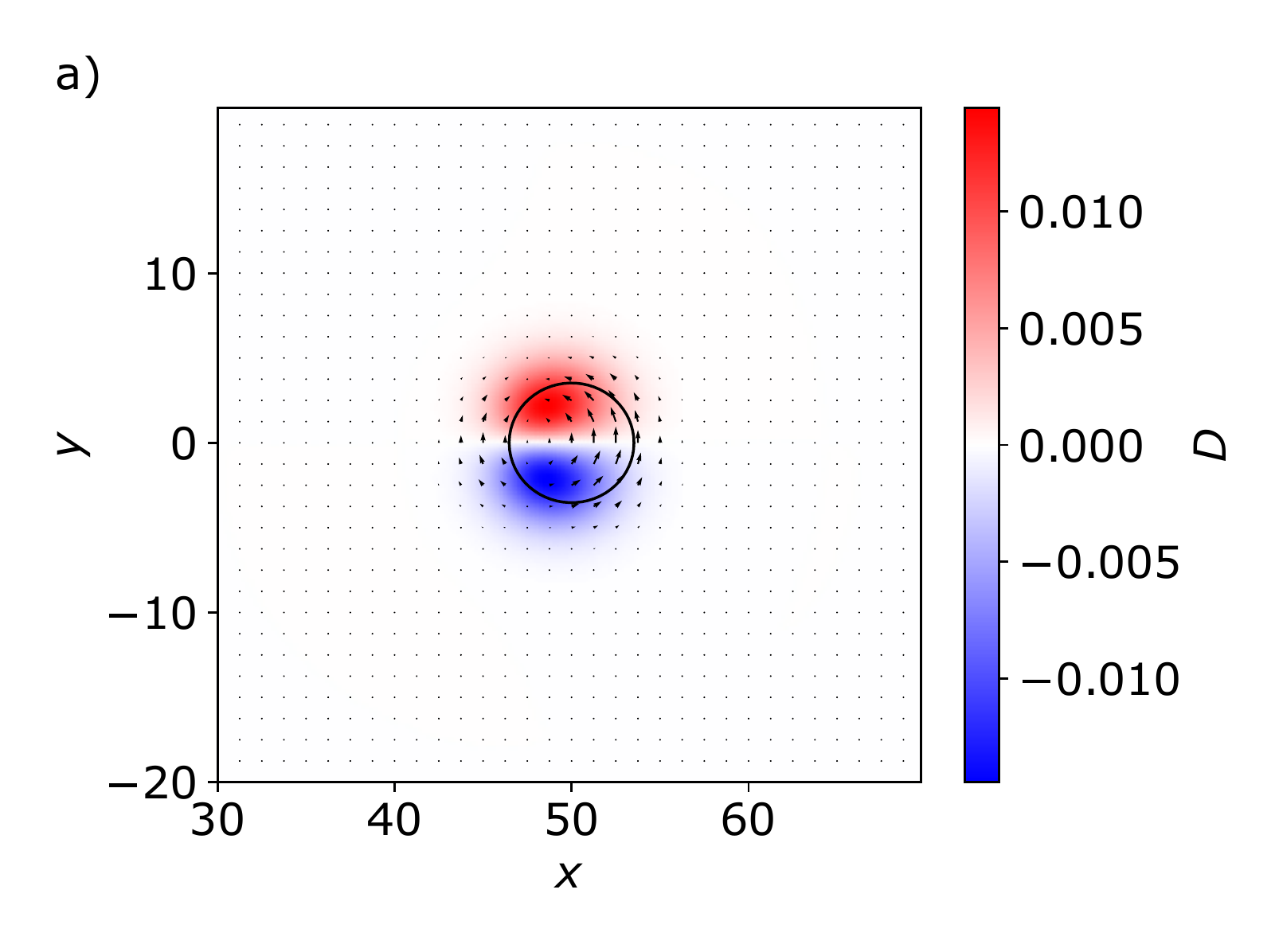}
     \includegraphics[width=0.43\textwidth]{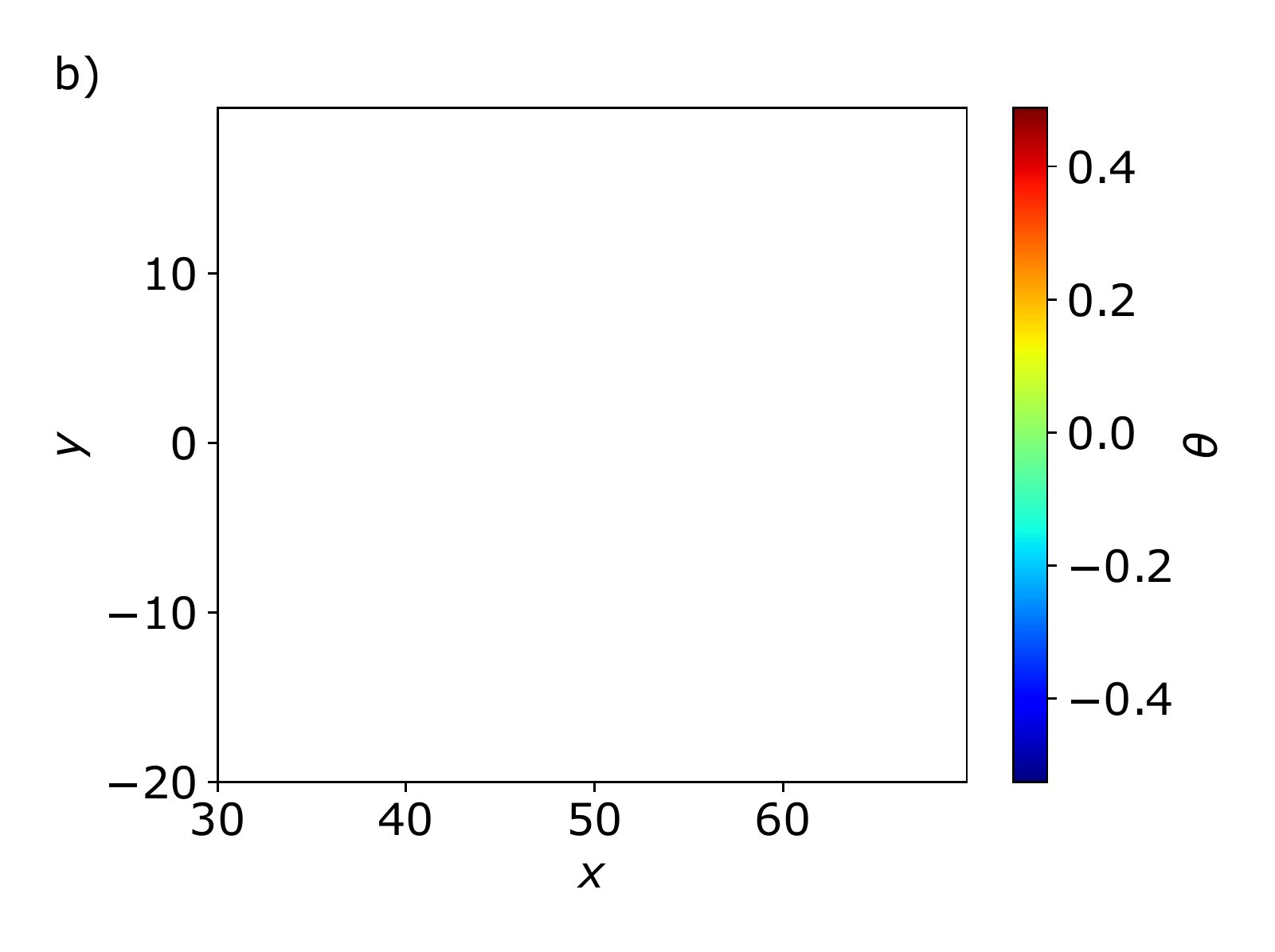}
    \includegraphics[width=0.43\textwidth]{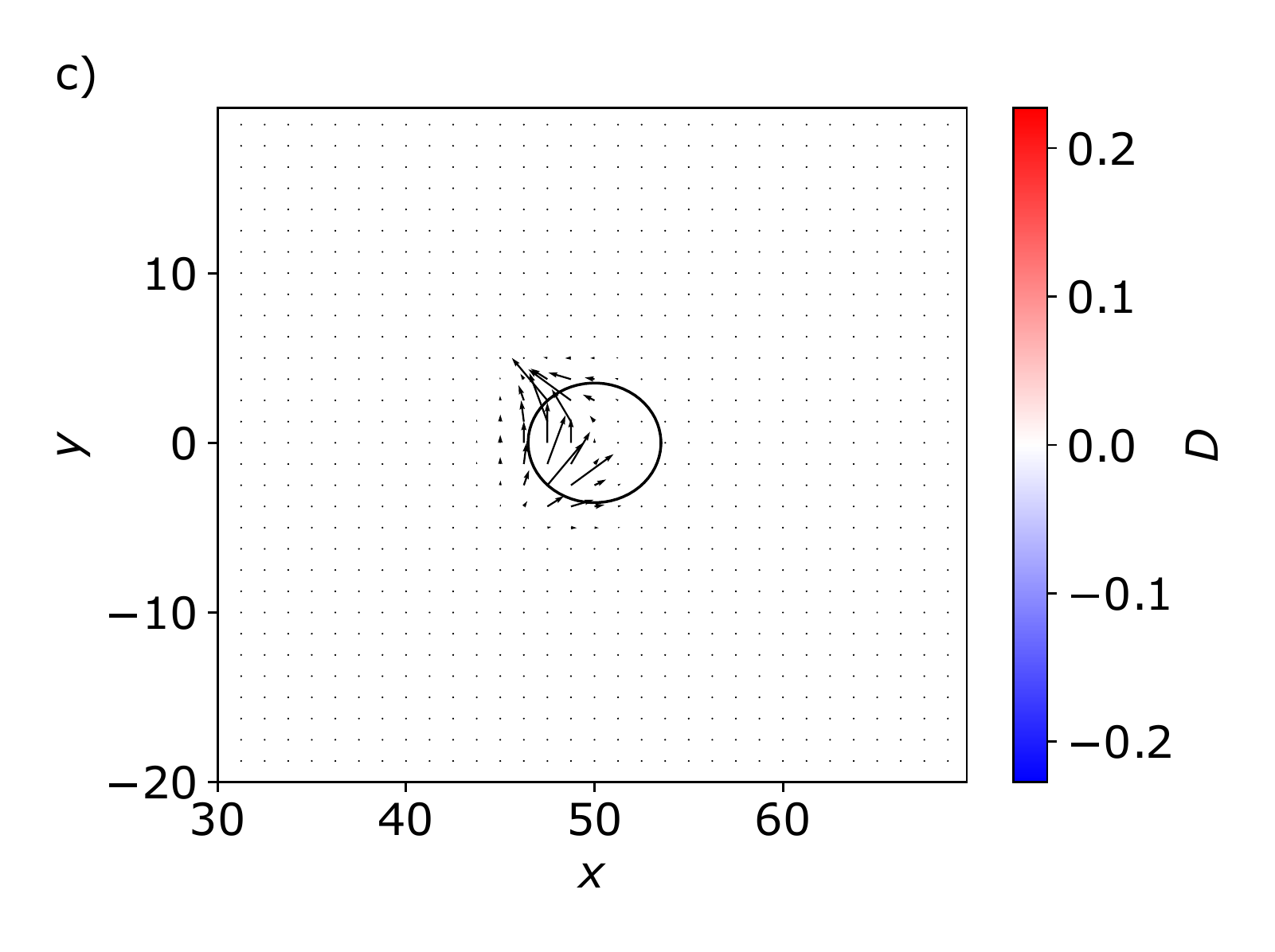} 
     \includegraphics[width=0.43\textwidth]{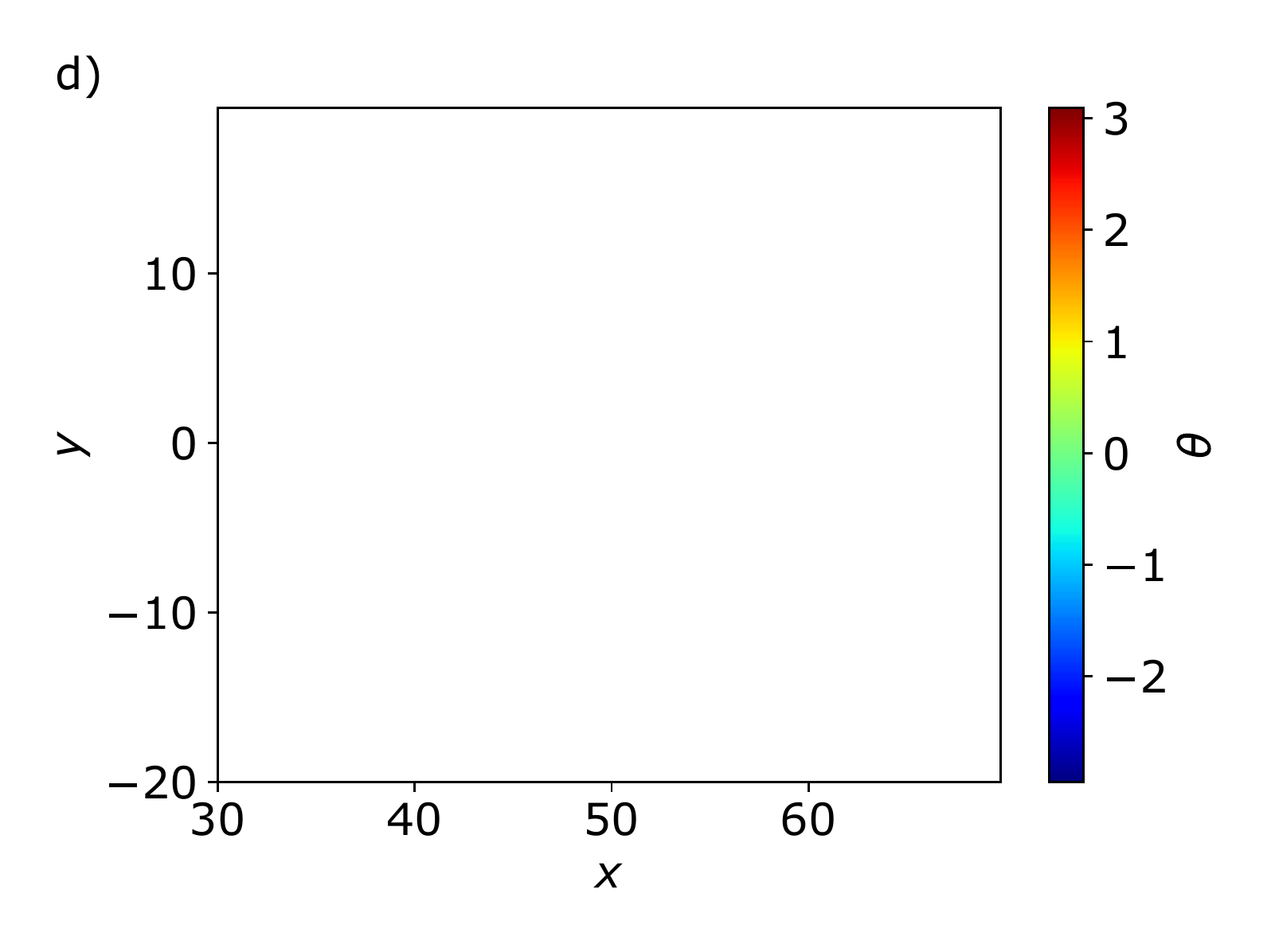}
    \caption{D-field with defect current (a,c) and condensate phase (b,d) around the soft potential during a dipole nucleation. Before the phase slips formed, $t=5$ (a,b), the D-field shows two smeared out "cores" inside the potential. These cores become denser and detach from the potential after the phase slips nucleated at $t=30$ (c,d). Stirring velocity is $V_0=0.4$ and the potential strength is $U_0 =0.8$.     }
    \label{fig:soft_pot}
\end{figure*}

The regime with $U_0> 1$ (in units of the chemical potential) corresponds to a hard potential whereby the condensate density almost vanishes inside the obstacle corresponding to the homogeneous boundary condition imposed at an impenetrable boundary. The opposite regime where $U_0 < 1$ is attributed to softly-indenting potentials where the condensate density is slightly depleted and phase coherence is preserved. We will show that these two limit cases have a different signature on the onset of vortex nucleation and shedding. This is apparent in Figs.~(\ref{fig:hard_pot}) and (\ref{fig:soft_pot}) that show the generalized vorticity $D$-field, its current and the phase of the wavefunction around a hard ($U_0=10$) and soft ($U_0=0.8$) potential, respectively. 

\begin{figure}[ht]
    \centering
    \includegraphics[width=0.43\textwidth]{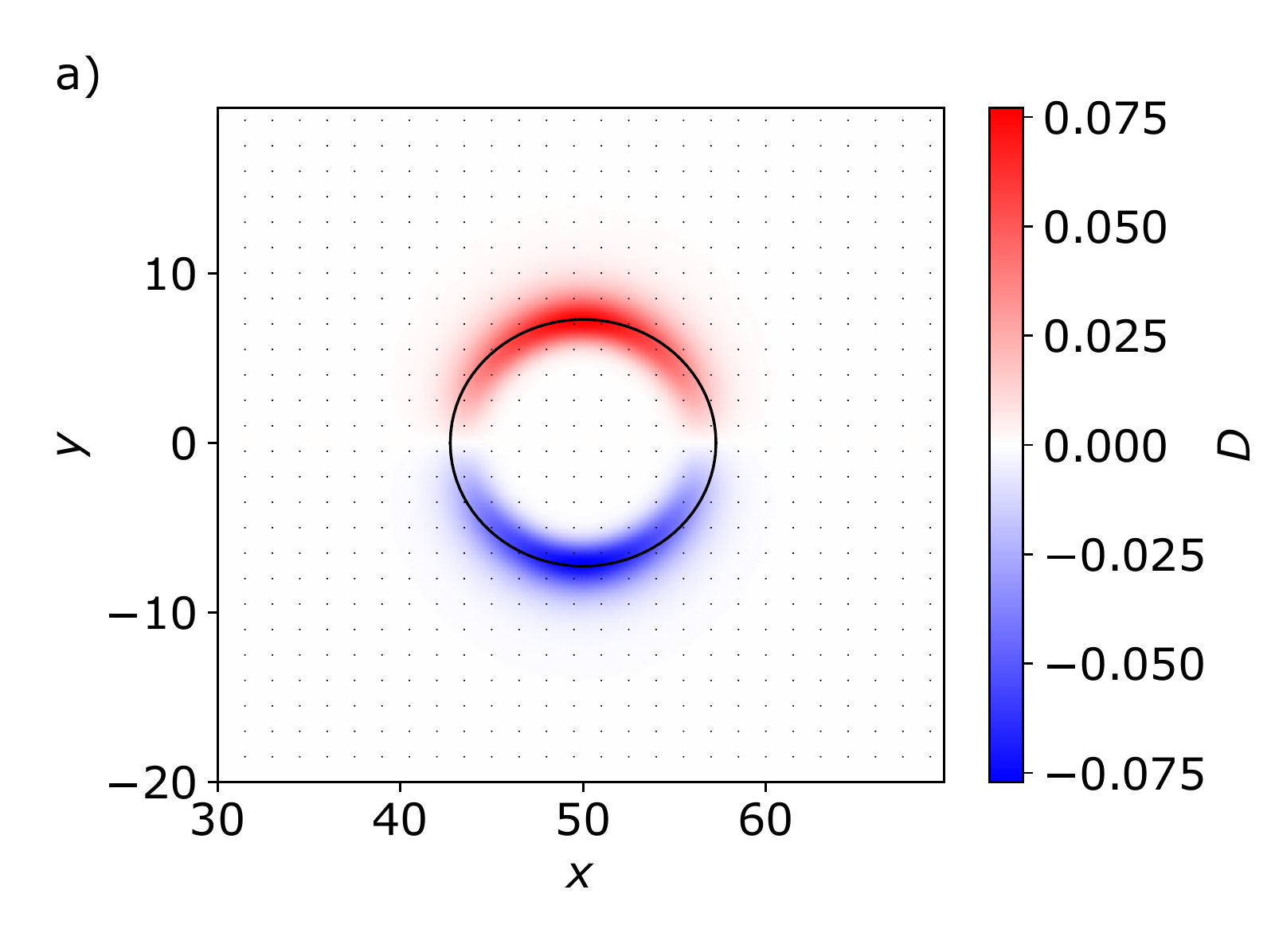}
    \includegraphics[width=0.43\textwidth]{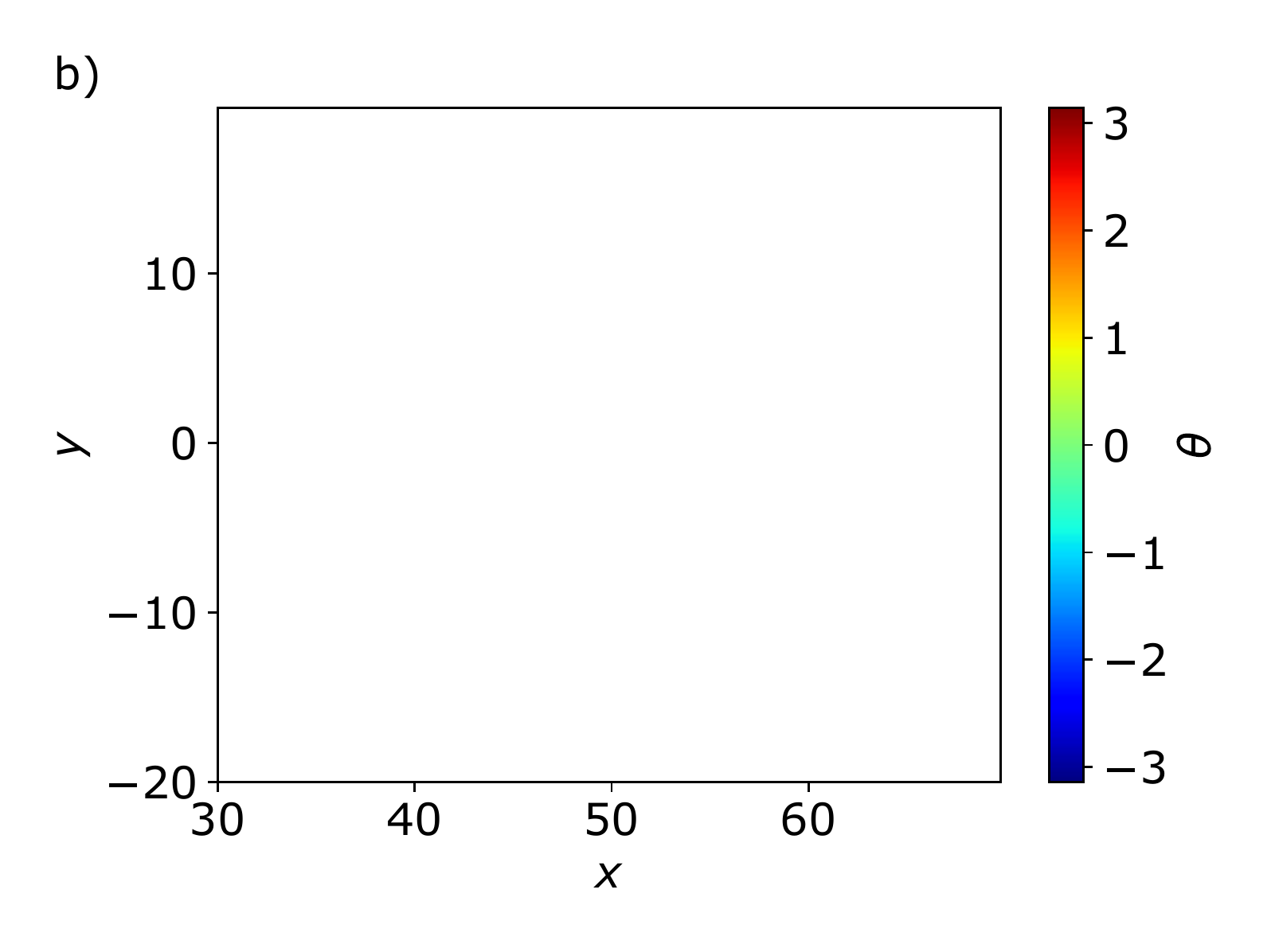}
    \caption{a) $D$-field forming a stable rim around the obstacle edge and accompanied by a vanishing generalized vorticity current. b) Stable ghost vortices as steady-state phase slips inside the obstacle at time $t=1800$. 
    }
    \label{fig:Hard_ghost}
\end{figure}

For the hard potential, phase slips develop inside the potential where the condensate density is close to zero as shown in Figure~\ref{fig:hard_pot}b). 
This is signaled by the concentration of the generalized vorticity around the edge of the potential. At the onset of nucleation and shedding, the dipole of phase slips separates and drifts toward the edge of the potential localizing the $D$-field into a vortex dipole with well-defined cores as shown in Figure~\ref{fig:hard_pot} c)-d). In Ref.~\cite{kunimi2015metastability}, it was shown that there are also stationary solutions of a dipole of phase slips, which were termed as ghost vortices because form inside the potential without ever nucleating vortices that are shed. A stationary configuration of such dipole of ghost vortices is also depicted in Figure~\ref{fig:Hard_ghost}, where we clearly see that phase slips reside in the middle of the potential while the generalized vorticity concentrates in a uniform halo with alternate sign in each half domain, and therefore the generalized vorticity current vanishes. 

For the soft potential, the precursory pattern and the onset to nucleation are different. Here, the $D$-field starts out as two smeared out regions of generalized vorticity with opposite sign and spanning the stirring potential accompanied by smooth phase gradients, i.e. no phase slips, as shown in Figure~\ref{fig:soft_pot} a) and b). At the onset of vortex nucleation at or above a critical speed $V_{0c}$,
a dipole of phase slips develops inside the potential while the generalize vorticity localises into two regions corresponding to two phase slips.
During the creation of the dipole, the generalised vorticity current connects the developing cores, which is due to $D$ being a conserved quantity.
When the vortex cores are formed, but the dipole is still under the potential, the vorticity current picks up the direction that the vortex dipole moves out of the potential as shown in Figure.~\ref{fig:soft_pot} c).
Just below the critical velocity $V_{0c}$, we see the formation of stable smeared out cores in the generalized vorticity, but without containing any phase slips. This is because the soft potential is not able to trap the phase slips inside it, thus any phase slip leads to nucleation and shedding of vortex dipoles. The transition from soft to hard potential obstacle occurs around $U_0 \sim 1$, where the potential is able to nucleate vortices and pin them underneath itself without shedding them into the bulk~\cite{kwak2022minimum}.

We have considered the homogeneous vortex nucleation and shedding away from a stirring potential in a uniform condensate. Because of the symmetry of the initial configuration and absence of noise, only dipoles are being nucleated and shed. However, a small noise added to the uniform condensate wavefunction breaks the symmetry of the initial state and may lead to vortex nucleation beyond simple dipoles~\cite{reeves2015identifying,sasaki2010benard}. The nucleation event remains symmetrical through the formation of dipoles of phase slips inside the potential. However, the shedding can become more irregular depending on the noise amplitude and stirring velocity.

\begin{figure}[ht]
    \centering
    \includegraphics[width=.43\textwidth]{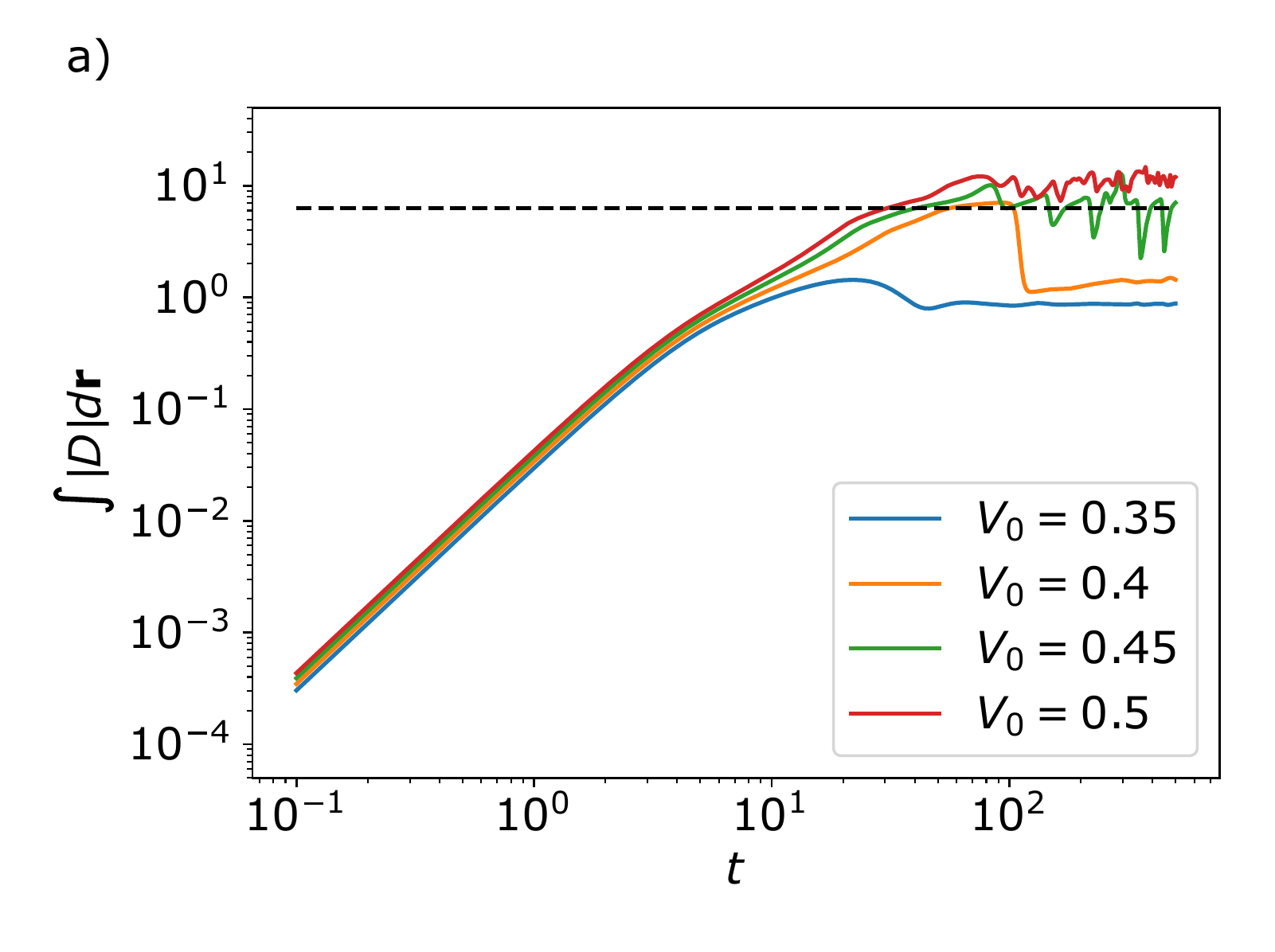}
    \includegraphics[width=.43\textwidth]{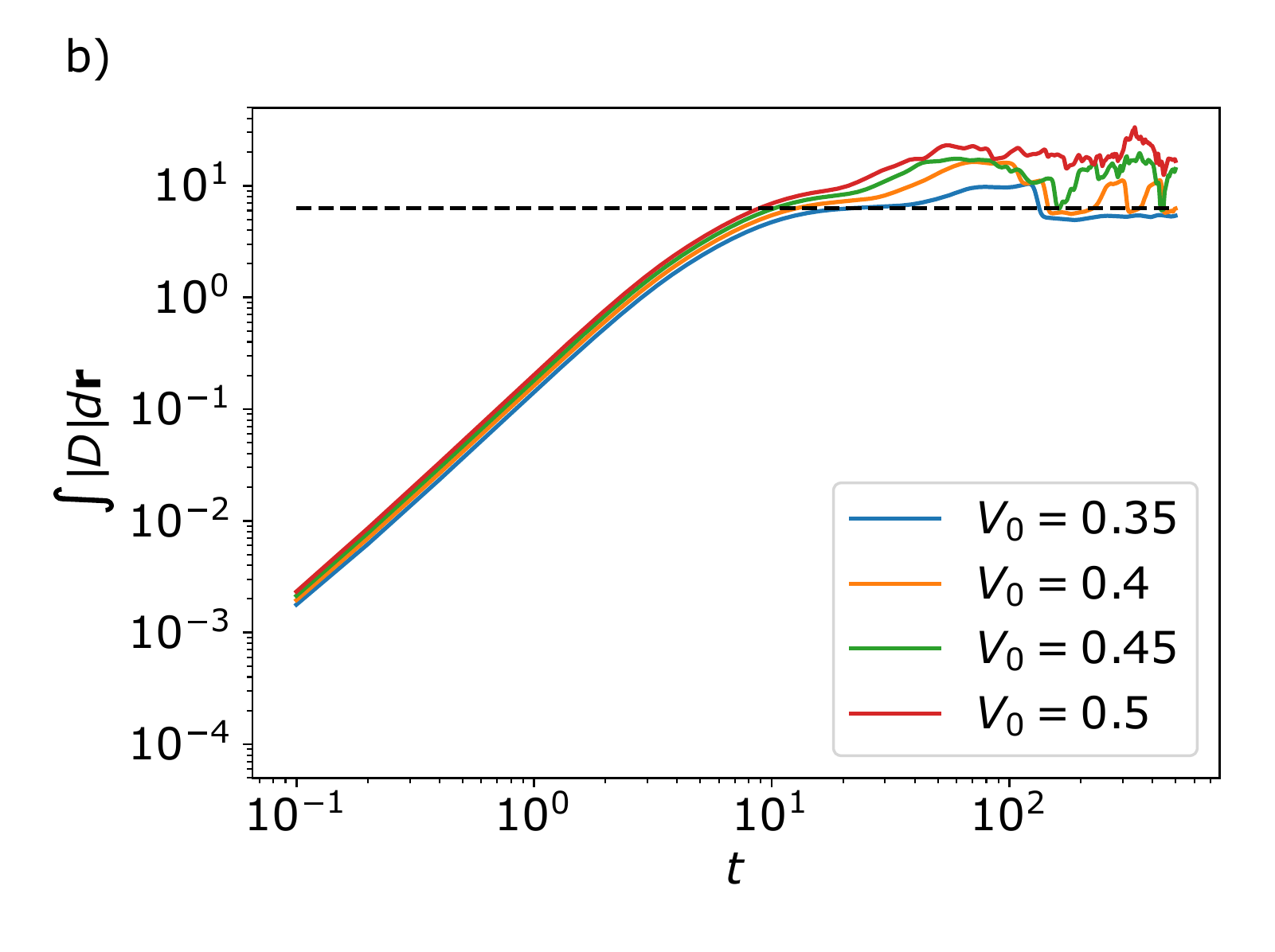}
    \caption{Net generalized vorticity for soft potential $U_0 = 0.8$ (a) and hard potential $U_0=10$ (b) for different $V_0$ close to the critical velocity. The integral is preformed over the area around the potential shown in Figs.~\ref{fig:hard_pot} and \ref{fig:soft_pot}. The doted line is the expected value corresponding to a single vortex dipole.  }
    \label{fig:V_0=0.8}
\end{figure}

To get a more quantitative measure of the nucleation event, we use the spatial average of the magnitude of the generalized vorticity $|D|$ as a proxy to the total number of vortices as shown in Eq.~(\ref{eq:abs_D_int}). The deviation from the theoretical prediction corresponding to a uniform superfluid punctuated by well-separated vortices informs us about the presence of additional density heterogeneities due to compressible modes or induced by the obstacle potential as discussed earlier and shown in Figs.~\ref{fig:hard_pot},~\ref{fig:Hard_ghost} and~\ref{fig:soft_pot}. 
In Figure~\ref{fig:V_0=0.8}, we have plotted this global measure as a function of time for a soft versus a hard potential, and for different stirring velocities. The integration domain is a square surrounding the obstacle and of size $l = 40$, i.e the same domain that is shown in Figs. \ref{fig:hard_pot} - \ref{fig:soft_pot}. 
We notice that there is an initial power-law growth $\omega(t) = \int_A d^2\mathbf r |D| \sim t^\beta$, $\beta\approx 2$ in both the soft and hard obstacle in the early times where the phase slips develop.  

Beyond this regime, we observe several time evolutions, depending on the stirring velocity. When $V_0<V_{0c}$, the global vorticity increases up to a peak value and then drops to a plateau with a slightly lower value which corresponds to the regime where the smeared out vorticity is formed around the potential in the absence of any phase slips or vortex shedding. This occurs for a soft potential as illustrated in Figure~\ref{fig:D-field}a). For $V_0\sim V_{0C}$ corresponding to the critical velocity for the nucleation of a single vortex dipole (in the absence of noise), the global vorticity reaches above the $2\pi$ threshold and signals the presence of a dipole which happens both for soft and hard potentials as shown in Figure~\ref{fig:D-field} a) and b). Of course, once the nucleated vortex dipole drifts out of the integration domain, the value of the integral drops and shows only the contribution of the vorticity halos around the potential. Finally, if $V_0>V_{0v}$, dipole shedding is a recurrent event highlighted by fluctuations of the global vorticity measure as shown in  Figure~\ref{fig:D-field}a) and b).

\section{Vortex kinematics}\label{sec:kinematics}
Using the Halperin-Mazenko formalism presented in Sec.~\ref{sec:MH_method}, we now derive the velocity of a vortex in the presence of both background superfluid flow and condensate density heterogeneities. To begin with, we consider a uniform condensate punctuated at the origin with a vortex. We expand the profile of the condensate wavefunction close to the defect core as $\psi_0 \approx re^{iq\theta}$ where $r$ is the distance to the vortex with circulation $q= \pm 1$ and  $r\ll 1$. For a stationary vortex, $\partial_t \psi_0 = 0$. We now introduce only a smooth phase perturbation $\psi(\mathbf r,t) = \psi_0(\mathbf r) e^{i\phi(\mathbf r,t)}$ which accounts for the net superfluid flow at the vortex position. For now, we consider that the background condensate density is constant. Near the defect core, the time evolution of the wavefunction is dominated by the kinetic term, 
\begin{equation}
    \partial_t \psi|_{\mathbf r= 0} \approx (i+\gamma)\frac{1}{2}\nabla^2\psi =  (-1+i\gamma)\nabla\psi_0 \cdot \nabla \phi e^{i \phi}. 
\end{equation}
The generalized vorticity $D$ is determined by the steady-state vortex wavefunction as 
\begin{equation}
    D = \frac{1}{2i}\epsilon_{ij}\partial_i\psi^*\partial_j\psi = \frac{1}{2i}\epsilon_{ij}\partial_i\psi_0^*\partial_j \psi_0.
\end{equation}
Thus, evaluating the vorticity current using the near-vortex evolution of the condensate wavefunction, we arrive at the following expression 
\begin{equation}
    J_i^{D} = \epsilon_{ij}[ -\epsilon_{jk}D + \gamma q \epsilon_{kl}\epsilon_{jk}D)]\partial_k\phi.
\end{equation}
which together with Eq.~(\ref{eq:vortex_velocity}), implies that the vortex velocity is determined by the phase gradients, hence the superfluid flow passing through its core 
\begin{equation}
    v_i = \left(\partial_i \phi + \gamma q \epsilon_{ij}\partial_j\phi\right)_{\mathbf r= 0}.
\end{equation}
which is the basic overdamped vortex dynamics in the point vortex model~\cite{kim2016role,skaugen2017origin}. However, this model does not include the effect of condensate density disturbances due to the presence of trapping or stirring potentials. We can apply the same method to compute the contribution of density variations to the vortex velocity. For this, the wavefunction is perturbed both in magnitude and phase: $\psi = \psi_0 e^{\lambda + i\phi}$, where $\phi$ and $\lambda$ are smooth real fields ~\cite{mazenko2001defect,groszek2018motion}. 
The generalized vorticity $D$-field acquires an additional contribution from the density perturbations and is given by  
\begin{equation}
    D = e^{2\lambda} \frac{1}{2i}\epsilon_{ij} \partial_i\psi_0 \partial_j \psi_0.
\end{equation}
The corresponding vortex velocity becomes 
\begin{widetext}
\begin{equation}
    v_{k} = 2i\frac{\textrm{Im}(\epsilon_{ij} \partial_j\psi_0 \partial_k \psi_0^*)[\partial_k \phi -\gamma\partial_k\lambda] + \textrm{Re}(\epsilon_{ij}\partial_j \psi_0 \partial_k \psi^*)[\partial_k\lambda + \gamma \partial_k\phi]}{\epsilon_{ij}\partial_i \psi_0^* \partial_j \psi_0}
\end{equation}
\end{widetext}
By using $i\partial_k \psi_0= q \epsilon_{kl} \partial_l\psi_0$, this reduces to a point vortex dynamics  
\begin{equation}\label{eq:point_vortex}
    v_{i} = \left(\partial_i \phi -\gamma\partial_i\lambda + \gamma q \epsilon_{ij} \partial_j \phi + q \epsilon_{ij} \partial_j \lambda\right)_{\mathbf r= \mathbf{0}}
\end{equation}
which is similar to the expressions obtained in Ref.~\cite{mazenko2001defect,groszek2018motion}.
In the later they discuss how to employ this expression, in the limit of no thermal dissipation $\gamma=0$, on point vortices in a harmonic trap. The orbiting of vortices is precisely determined by last term in Eq.~(\ref{eq:point_vortex}) due to the spatial profile of the condensate density.   
The effect of thermal drag is that it makes oppositely charged vortices attract each other according to the third term in Eq.~(\ref{eq:point_vortex}). Also, vortices move down gradients in the background condensate density as given by the second term. 
The later effect is shown in Figure~\ref{fig:mesh_vs_mazenco} a) where the vortex acquires a non-zero radial velocity due to non-zero $\gamma$ and inhomogeneous profile of the condensate density induced by the harmonic trap.


\section{Discussion and conclusions} \label{sec:discussion}
In summary, we have applied the Halperin-Mazenko formalism to characterise the nucleation and dynamics of vortices in a stirred Bose-Einstein condensate. This formalism allows to us introduce a smooth generalized vorticity $D$-field as a topologically conserved quantity and an associated vorticity current which tracks all the disturbances in the condensate both singular (vortices) and non-singular (phonons or smooth disturbances induced by external potentials). When the uniform condensate is stirred by a Gaussian potential, the onset to vortex nucleation is predicted by the precursory pattern formations in the generalized vorticity field which depend on the permeability of the obstacle. Namely, the $D$-field shows the formation of two diffusive clouds around the soft potential before any phase slips develop. These diffusive clouds becomes localised into two well-defined cores during the formation of a dipole of the phase slips inside these regions which corresponds to the onset of vortex nucleation. While the onset of nucleation is consumed inside the potential, the actual nucleation is manifested into the condensate by the shedding of the vortex dipole.
For the hard potential, initial diffusive clouds of the generalized vorticity are also formed around the edge of the potential while phase slips develop inside it.  
At the onset of nucleation, a dipole of phase slips moves to the edge and localises the $D$-field into a dipole of vortex cores. In both cases, the nucleation of vortices is a gradual process consumed inside the potential. 
The generalized vorticity field is related to a current density through a conservation equation. The ratio of the defect density current and the defect density  determines the vortex velocity. We have used this generic formula to derive known expressions for the vortex velocity corresponding to a weekly-perturbed condensate.  It is worth noting that the Halperin-Mazenko formalism may be extended also to analysing experimental data and identifying different types of condensate disturbances.

\begin{acknowledgements}
We are grateful to Vidar Skogvoll, Marco Salvalaglio and Jorge Vi\~nals for many fruitful discussions. 
\end{acknowledgements}

 \bibliographystyle{apsrev4-2}
 \bibliography{references}

\end{document}